\documentstyle[amstex,amssymb,preprint,aps,epsfig]{revtex}
\newcommand{\1}{{\Omega_{\rm M} }}
\newcommand{\2}{{\Omega_{{\rm M0}} }}

\newcommand{\beq}{\begin{equation}}
\newcommand{\eeq}{\end{equation}}
\newcommand{\lab}{\label}

\begin{document}
\draft
\title{The peculiar velocity field in a quintessence model}
\author{Claudio Rubano and Mauro Sereno}
\address{Dipartimento di Scienze Fisiche, Universit\`{a} Federico II\\
and \\ Istituto Nazionale di Fisica Nucleare, Sez. di Napoli,\\
Complesso Universitario di Monte S. Angelo,\\ Via Cintia, Ed. G,
I-80126 Napoli, Italy\\ }
\maketitle

\begin{abstract}
We investigate the evolution of matter density perturbations and some
properties of the peculiar velocity field for a special class of
exponential potentials in a scalar field model for quintessence, for
which a general exact solution is known. The data from the 2-degree
Field Galaxy Redshift Survey (2dFGRS) suggest a value of the today
pressureless matter density $\Omega_{{\rm M0}}=0.18 {\pm} 0.05$.

KEYWORDS: cosmology: theory - cosmology: quintessence - large-scale
structure of Universe
\end{abstract}

\narrowtext
\section{Introduction}

In the last few years, a new class of cosmological models has been proposed. The more
usual cosmological constant $\Lambda$-term is replaced with a dynamical, time dependent
component, the quintessence or dark energy, added together with baryons, cold dark
matter (CDM), photons and neutrinos. The equation of state of the dark energy is given
by $w_Q \equiv \rho_Q /p_Q$, $\rho_Q$ and $p_Q$ being, respectively, the pressure and
energy density; since $-1 \leq w_Q <0$, it still implies a negative contribution to the
total pressure of the cosmic fluid. When $w_Q
=-1$, we recover a constant $\Lambda$-term. One of the more interesting physical
realizations of the quintessence is a cosmic scalar field minimally coupled with the
usual matter action \cite{pe+ra88,cal+al98}. Such a field induces the repulsive
gravitational force dynamically, allowing to explain the accelerated expansion of our
Universe, as discovered with observations of SuperNovae (SNe) of type Ia
\cite{per+al99,rie+al98}.

Together with an accelerated expansion, another experimental fact is the strong
evidence of a spatially flat Universe \cite{deb+al00}: quintessence could be
responsible for the missing energy in a flat Universe with a subcritical matter
density.

Since quintessence drives the cosmological expansion at late times, it also influences
the growth of structures arisen from gravitational instability. Quintessence clusters
gravitationally at very large scales ($\stackrel{>}{\sim} 100$ Mpc), leaving an imprint
on the microwave backgroud anisotropy \cite{cal+al98}; at small scales, fluctuations in
the dark energy are damped and do not enter in the evolution equation for the
perturbations in the pressureless matter \cite{ma+al99}. At the scale we are
considering in the following, quintessence behaves as a smooth component: it does not
partecipate directly in cluster formation, but it only alters the background cosmic
evolution.

In this work, we study the evolution of density perturbations for a general exact
solution for quintessence model, found by one of us together with P. Scudellaro
(hereafter the RS model) \cite{ru+sc01}. In Section 2, we present the basic equations
of the quintessence models examined in the paper. In Section 3, the linear
perturbations equation is solved for the RS potential and the properties of the
peculiar velocity field are discussed; in Section 4, the 2dFGRS data are considered.
Section 5 is devoted to some final considerations.

\section{Model description}

We investigate spatially flat, homogeneous and isotropic cosmologies filled with two
non-interacting components: pressureless matter (dust) and a scalar field $\varphi$,
minimally coupled with gravity. We consider the RS potential,
\begin{equation}
\lab{scal1}
V(\varphi) \propto \exp\left\{ -\sigma \varphi\right\},
\end{equation}
where $\sigma^2 \equiv 12 \pi {\rm G}/{\rm c}^2$. We refer to Rubano
\& Scudellaro \cite{ru+sc01} and Pavlov et al. \cite{pav+al02}
for details and a discussion of the general exact solution of the
field equations and we present here only what is needed for the
present work.\footnote{In particular, in Rubano \& Scudellaro
\cite{ru+sc01}, it is explained why this potential works, on the
contrary of a quite diffuse opinion.} It is
\begin{equation}
\lab{scal2}
H=\frac{2(1+2\tau^2)}{3t_s \tau (1+\tau^2)},
\end{equation}
and
\begin{equation}
\lab{scal3}
\1 =\frac{1+\tau^2}{(1+2\tau^2)^2},
\end{equation}
where $\tau \equiv t/t_s$ is a dimensionless time; $t_s$ is a time scale of the order
of the age of the Universe; $H$ is the time dependent Hubble parameter; $\1$ is the
pressureless matter density ($\rho_M$) in units of the critical density $\rho_{{\rm
cr}}=3H^2 /8\pi{\rm G}$.

The relation between the dimensionless time $\tau$ and the redshift $z$ is given by
\begin{equation}
\lab{scal4}
(1+z)^3=\frac{\tau_0 (1+\tau_0^2)}{\tau (1+\tau^2)},
\end{equation}
where $\tau_0$ is the present value of $\tau$. This very simple cosmological model has
two free parameters, $t_s$ and $\tau_0$, or, equivalently, $H_0$, the today value of
the Hubble parameter, and $\2$, the present value of the pressureless matter density.
As can be seen from Eqs.~(\ref{scal3},\ref{scal4}), $\2$ depends only on $\tau_0$

We shall compare the results for the above Universe with a flat model with constant
equation of state $w_Q$ (as well known, this includes the $\Lambda$CDM cosmologies for
$w_Q=-1$). In this case, the redshift dependent Hubble parameter is
\begin{equation}
\lab{scal5}
H(z)=H_0\sqrt{\2(1+z)^3+ (1-\2)(1+z)^{3(1+w_Q)}}.
\end{equation}
This class of models has three free parameters, $\2$, $H_0$ and $w_Q$.

\section{Perturbation growth and peculiar velocity}

For perturbations inside the horizon, the equation describing the
evolution of the CDM component density contrast, $\delta_M
\equiv \delta \rho_M /\rho_M$ (for unclustered quintessence) is
\cite{peeb80,ma+al99}
\begin{equation}
\lab{grow1}
\ddot{\delta}_M+2H(t)\dot{\delta}_M-4\pi{\rm G }\rho_M \delta_M=0,
\end{equation}
where the dot means derivative with respect to comoving time. In Eq.~(\ref{grow1}), the
relative contribution of dark energy to the energy budget enters into the expansion
rate $H$. We shall consider Eq.~(\ref{grow1}) in the matter dominated era, when the
radiation contribution is really negligible.

With a RS potential, Eq.~(\ref{grow1}) reduces to
\begin{equation}
\lab{grow2}
\frac{d^2\delta_M}{d\tau^2}+
\frac{4}{3}\frac{(1+2\tau^2)}{\tau (1+\tau^2)}\frac{d\delta_M}{d\tau}-
\frac{2}{3}\frac{1}{\tau^2 (1+\tau^2)}\delta_M=0.
\end{equation}
Equation~(\ref{grow2}) has two linearly independent solutions, the growing mode
$\delta_+$ and the decreasing mode $\delta_-$, which can be expressed in terms of the
hypergeometric function of second type $\ _2F_1$. We get
\begin{equation}
\lab{grow3}
\delta_- \propto \frac{1}{\tau}
\ _2F_1 \left[ -\frac{1}{2},\frac{1}{3},\frac{1}{6}; -\tau^2 \right],
\end{equation}
and
\begin{equation}
\lab{grow4}
\delta_+ \propto \tau^{2/3}
\ _2F_1 \left[ \frac{1}{3},\frac{7}{6},\frac{11}{6}; -\tau^2 \right].
\end{equation}
By comparing Eqs.~(\ref{grow3},\ref{grow4}) with Eq.~(\ref{scal4}), we observe that the
density contrast, as a function of the redshift, depends on only one parameter,
$\tau_0$, i.e. $\2$. For $\tau \ll 1$, Eqs.~(\ref{grow3},\ref{grow4}) reduce to the
standard results
\begin{equation}
\lab{grow5}
\delta_- \propto \frac{1}{\tau},
\end{equation}
and
\begin{equation}
\lab{grow6}
\delta_+ \propto \tau^{2/3}.
\end{equation}
For the evolution of the density contrast for quintessence with
constant equation of state, we refer to Silveira and Waga
\cite{si+wa94}.

\begin{figure}
        \epsfxsize=10cm
        \centerline{\epsffile{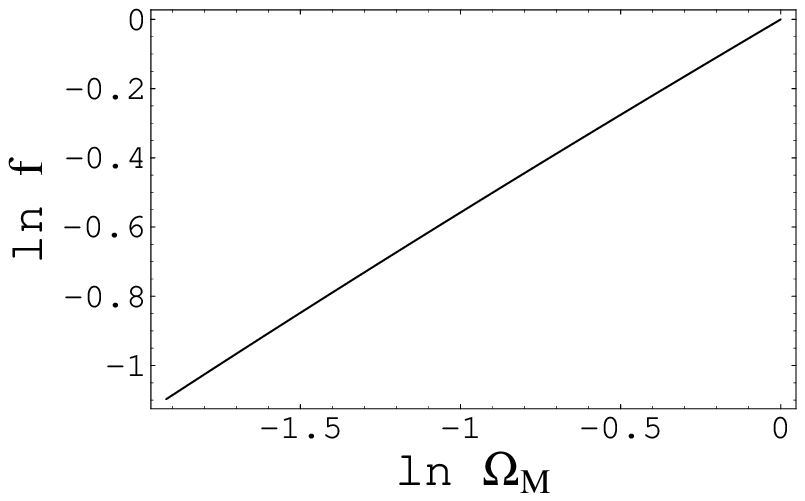}}
        \caption{\small $\ln f$ versus $\ln \1$ in the RS model.}
        \label{ln-f-ln-omega}
\end{figure}

The linear theory relates the peculiar velocity field ${\bf v}$ and the density
contrast by \cite{peeb80,padm93}
\begin{equation}
\lab{grow7}
{\bf v} ({\bf x})= H_0 \frac{f}{4\pi} \int \delta_M ({\bf y}) \frac{{\bf x}-{\bf
y}}{\left| {\bf x}-{\bf y} \right|^3} d^3 {\bf y},
\end{equation}
where the growth index $f$ is defined as
\begin{equation}
\lab{grow8}
f \equiv \frac{d \ln \delta_M}{d \ln a};
\end{equation}
$a$ is the scale factor, $a =a_0/(1+z)$, and $a_0$ is its present value. For a RS
potential, we get
\begin{equation}
\lab{grow9}
f = \frac{2}{3}\frac{(1+2\tau^2)}{\tau (1+\tau^2)}\frac{d \ln \delta_M}{d \tau}.
\end{equation}
The growth index is usually approximated by $f \simeq \1^\alpha$.
For $\Lambda$CDM models, $\alpha \simeq 0.55$; if $w_Q = -2/3$, $\alpha \simeq 0.56$;
if $w_Q = -1/3$, $\alpha \simeq 0.57$ \cite{si+wa94,wa+st98}. In
Fig.~(\ref{ln-f-ln-omega}), we show the logarithm of $f$ as a function of the logarithm
of $\1$ for the RS potential. The value $\alpha
\simeq 0.57$ provides a good approximation to the model.

\begin{figure}
        \epsfxsize=10cm
        \centerline{\epsffile{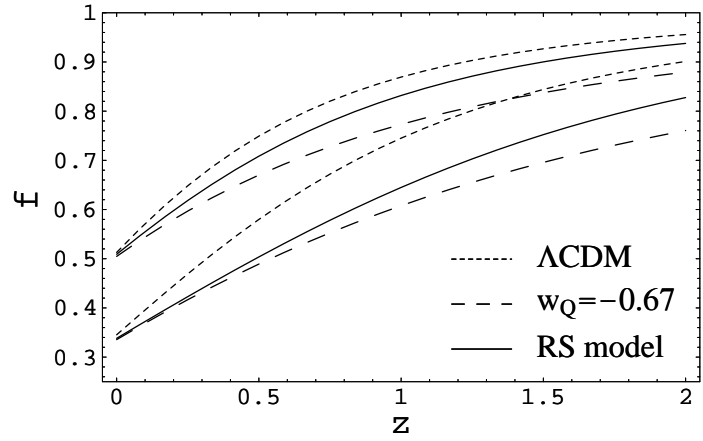}}
        \caption{\small The growth index as a function of the redshift for two values of $\2$.
        The upper curves are for $\2 =0.3$, the lower curves for $\2 =0.15$. The dashed,
        long-dashed and full lines correspond, respectively, to $\Lambda$CDM cosmologies,
        models with $w_Q=-2/3$ and the RS model.}
        \label{f-z}
\end{figure}

In order to compare the various models, we have to stress that the ``right" value for
$\2$ depends on the model. It is the result of a best fit prodecure on experimental
data, i.e. from SN Ia observations. The well known value $\2 \simeq 0.3$ refers only to
the $\Lambda$CDM model. With the same procedure applied to SNe Ia data, the RS
potential gives $\2 =0.15^{+0.15}_{-0.03}$ \cite{pav+al02}. Due to the large errors,
the two results are still marginally compatible, so the value $\2 \simeq 0.3$ holds for
the RS model too. In the case of a constant $w_Q \neq -1$, the situation is more
complicate, due to the degeneration between  $\2$ and $w_Q$ \cite{per+al99,ser+al01}.
In this preliminary study, we decided to consider only two situations: {\it i)} a
$\Lambda$CDM, a $w_Q \neq -1$ and a RS model with $\2 =0.3$; {\it ii)} the same three
models with $\2 =0.15$. As can be seen from Fig.~(\ref{f-z}), a large variation in $\2$
entails a remarkable difference in the growth index also at low redshifts. Even more
remarkable is the difference at large redshifts in the case of the same $\2$. This
suggest that with the deep surveys planned in the future, this approach could provide a
constraint independent of SNe Ia \cite{ze+de99}.

\begin{figure}
        \epsfxsize=10cm
        \centerline{\epsffile{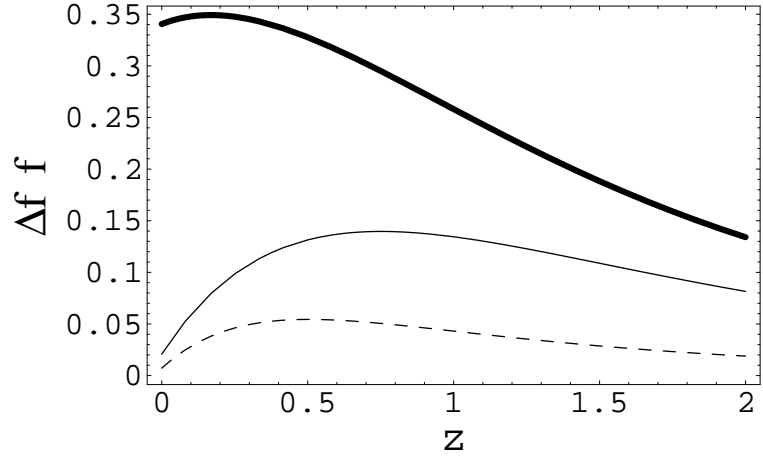}}
        \caption{\small Relative variations of the growth index in the RS model from the
        $\Lambda$CDM case as a function of the redshift. For the full thin line, $\2 =0.15$;
        for the dashed line, $\2 =0.3$. With the thick curve we indicate the variation of
        the RS model with $\2 =0.15$ from a $\Lambda$CDM model with $\2 =0.3$.}
        \label{deltaf-z}
\end{figure}

In Fig.~(\ref{deltaf-z}), we show the relative variations of RS models from
$\Lambda$CDM cosmologies. For models with the same $\2$, the relative variations are
quite small and decrease with increasing pressureless matter density. For $\2= 0.3$,
the relative variation is $\stackrel{<}{\sim} 5\%$; for $\2= 0.15$, it is
$\stackrel{<}{\sim} 15\%$.

\section{A first check with observations}
Low redshifts have been investigated by the 2dFGRS collaboration \cite{pea+al01}.
Peculiar velocity distorts the correlation function of galaxies observed in redshift
space \cite{kais87,hami92}. In linear theory, the characteristic quadrupole distortion
enables to measure the growth index from redshift galaxy catalogs. The 2dFGRS has
obtained redshifts for more than $141,000$ galaxies with an effective depth of $z=0.17
$. From a precise measurement of the clustering, a value of $\beta \equiv f/b = 0.43 {\pm}
0.07$ has been determined \cite{pea+al01}, where $b$ is the bias parameter, connecting
the relative density fluctuations of the galaxies and of the total mass. The bias
parameter has been measured by computing the bispectrum of the 2dFGRS; it is $b=1.04 {\pm}
0.11$ \cite{ver+al01}. Combining the measurements of $\beta$ and $b$, the growth index
at $z=0.17$ can be estimated: it is $f(z=0.17)=0.45{\pm} 0.06$ \cite{ver+al01}.
By using the RS potential, from this value of $f$ we can obtain an estimate of the
today pressureless matter density parameter. We get
\begin{equation}
\lab{grow10}
\2 =0.18 {\pm} 0.05.
\end{equation}
This value is significantly compatible with the estimate from the SN Ia data.

\section{Conclusions}
The present state of accuracy in observations does not allow to discriminate among the
illustrated alternatives.

In our opinion, the above discussion adds another argument in favour of the use of
exponential potentials in quintessence cosmology. The strong dependence of the growth
index on the model, at high redshifts, suggests that more stringent constraints can be
given in the future.

\section*{Acknowledgments}
This work has been in part financially sustained by the M.U.R.S.T.
grant PRIN2000 ``SIN.TE.SI.".

\end{document}